\documentclass[preprint,showpacs,preprintnumbers,amsmath,amssymb,amsfonts]{revtex4-2}
\usepackage{graphicx,latexsym,amssymb,amsmath,color,multirow,mathrsfs,booktabs,ifpdf}
\def\bm#1{\mbox{\boldmath $#1$}}
\def\nA{nucleon-nucleus\ }
\def\aA{$\alpha$-nucleus\ }
\def\AA{nucleus-nucleus\ }
\def\nPb{$n+^{208}$Pb\ }
\def\pPb{$p+^{208}$Pb\ }
\def\pCa{$p+^{40}$Ca\ }

\begin{document}
\title{Pauli nonlocality and the nucleon effective mass}
\author{Dao T. Khoa$^1$}
\author{Doan Thi Loan$^1$}
\author{Nguyen Hoang Phuc$^2$}
\affiliation{$^1$ Institute for Nuclear Science and Technology, VINATOM, 179 Hoang Quoc Viet, 
 Hanoi 122772, Vietnam. \\
 $^2$ Phenikaa Institute for Advanced Study (PIAS), Phenikaa University, Hanoi 12116, Vietnam.}
\begin{abstract}
{A study of the nucleon mean-field potential in nuclear matter (NM) is done within an extended 
Hartree-Fock (HF) formalism, using the CDM3Y6 density dependent version of the M3Y interaction 
which is associated with the nuclear incompressibility $K\simeq 252$ MeV. The momentum dependence 
of nucleon optical potential (OP)  in NM at the saturation density $\rho_0$ is shown to be due mainly 
to its exchange term up to $k\approx 2$ fm$^{-1}$, so that the Pauli nonlocality is expected to be the 
main origin of the nucleon effective mass at low momenta. Because nucleons in neutron-rich NM at 
$\rho\approx \rho_0$ are either weakly bound or unbound by the in-medium nucleon-nucleon interaction, 
the determination of the effective mass of nucleon scattered on targets with neutron excess at low energies 
should be of interest for the mean-field studies of neutron star matter.  For this purpose, the folding model 
is used to calculate the nonlocal nucleon OP for the optical model analysis of elastic nucleon scattering 
on $^{40,48}$Ca, $^{90}$Zr, and $^{208}$Pb targets at energies $E<50$ MeV, to probe the model reliability 
and validate the WKB local approximation to obtain the local folded nucleon OP. The nucleon effective 
mass $m^*$ is then carefully deduced from the momentum dependence of the local folded nucleon OP 
which is resulted from the Pauli nonlocality of the exchange term. The obtained $m^*$ values agree well 
with the nucleon effective mass given by the extended HF calculation of the single-particle potential in 
asymmetric NM. The neutron-proton effective mass splitting determined at $\rho\approx\rho_0$ from the central 
strength of the real folded nucleon OP for $^{48}$Ca, $^{90}$Zr, and $^{208}$Pb targets has been found 
to depend linearly on the neutron-proton asymmetry parameter as $m^*_{n-p}\approx (0.167\pm 0.018)\delta$, 
in a good agreement with the recent empirical constraints.} 
\end{abstract}
\date{\today}
\maketitle

\section{Introduction}
\label{intro} 
Over a wide range of the single-particle energies, the nucleon motion in medium is overwhelmingly 
governed by the nuclear mean field, known as the shell-model potential for bound states and optical 
potential (OP) for scattering states. The single-particle potential is also a key quantity in the mean-field 
studies of the equation of state (EOS) of nuclear matter (NM) as well as the structure of finite nuclei
 \cite{Mah85,Bal07}. The nucleon OP was widely studied in the Brueckner-Hartree-Fock (BHF) 
calculations of NM using free nucleon-nucleon (NN) interaction \cite{Bal07,Bom91,Zuo99,Zuo14}, 
and in the mean-field calculations of NM on the Hartree-Fock (HF) level using different choices of the 
effective NN interaction \cite{Kho93,Kho95,Xu10,Che12,Xu14,Loa15}. The mean-field description 
of nucleon OP in the NM limit provide an important physics input for the microscopic models 
of nucleon OP of finite nuclei, in particular, different versions of the folding model 
\cite{Loa15,Ken01,Kho02,Kho07,Minomo}.  

The microscopic studies of NM have repeatedly shown the impact by the Pauli blocking and  
increasing strength of higher-order NN correlations at high densities \cite{Bal07}. These in-medium 
effects are considered as the physics origin of the density dependence introduced into various 
effective NN interactions used in the nuclear structure and reaction studies. In the present work, 
we employ the CDM3Y6 density dependent version \cite{Kho97} of the M3Y interaction 
\cite{An83} which was used successfully in the HF studies of NM 
\cite{Kho93,Kho95,Kho96,Tha09,Loa11}, and in the folding model calculation of the nucleon 
and nucleus-nucleus OP at low and medium energies  
\cite{Kho97,Kho02,Kho07,Kho07r,Kho09,Kho14,Loa15,Kho16}. 
The CDM3Yn density dependence (n=1-6) of the original M3Y interaction \cite{An83} was 
first parametrized in Ref.~\cite{Kho97} to reproduce the saturation of symmetric NM on the HF level. 
The CDM3Y3 and CDM3Y6 versions were later modified to take into account the rearrangement term 
(RT) of the single-particle potential in NM into the folding calculation of the \nA and \AA OP 
\cite{Loa15,Kho16}, based on the Hugenholtz and van Hove (HvH) theorem \cite{HvH,Sat99} 
which is exact for all interacting Fermi systems independently of the interaction between fermions.  

When the antisymmetrization of the \nA system is taken explicitly into account, the exchange term 
of the folded \nA potential becomes {\em nonlocal} in coordinate space \cite{Ken01,Minomo}, and 
a nonlocal folding model of nucleon OP using the CDM3Yn interaction was suggested in Ref.~\cite{Loa20}, 
with the RT properly taken into account. The calculable $R$-matrix method \cite{desco, desco2} 
is used to solve the OM equation with the {\em nonlocal} kernel of the folded nucleon OP. This method 
was well tested in the OM analysis of elastic nucleon scattering at energies up to 40 MeV \cite{Loa18}, 
using the phenomenological nonlocal nucleon OP \cite{Perey,TPM,Lovel,Lovel2}.  

The momentum dependence of nucleon OP at the NM saturation density $\rho_0$ is due mainly 
to the exchange term up to $k\approx 2$ fm$^{-1}$,  and the antisymmetrization of the \nA system 
is, therefore, the main origin of the momentum dependence of nucleon OP at low momenta. 
As a result, it becomes possible to determine the nucleon effective mass from the momentum 
dependence of the folded nucleon OP of finite nuclei, based on the WKB local approximation, 
in the OM analysis of low-energy elastic nucleon scattering on target nuclei with neutron excess. 
The present work is our first attempt to determine the radial- and isospin dependence of the nucleon 
effective mass near the saturation density ($\rho\lesssim\rho_0$) from the radial strengths of the real 
folded nucleon OP. The obtained results for $m^*$ allowed us to determine the neutron-proton 
effective mass splitting $\Delta m^*$ at $\rho\approx\rho_0$ for $^{48}$Ca, $^{90}$Zr, and 
$^{208}$Pb targets, which is complementary to the $\Delta m^*$ values obtained in different 
mean-field calculations of asymmetric NM. 

The knowledge about the nucleon effective mass is of wide importance for different nuclear 
physics and nuclear astrophysics studies \cite{Hog83,Li13,XLi15,Li15,Li18}. In particular,
the direct link of the nucleon effective mass to the density dependence of nuclear symmetry 
energy \cite{Zuo14,Hor14,Ba08}, the liquid-gas phase transition in the neutron-rich NM, and 
the temperature profile of hot protoneutron stars \cite{Tan16} and neutrino emission 
therefrom \cite{Bal14}. 

\section{Single-particle potential in nuclear matter}
\label{sec1} 
\begin{figure}[bht] 
\includegraphics[width=0.8\textwidth]{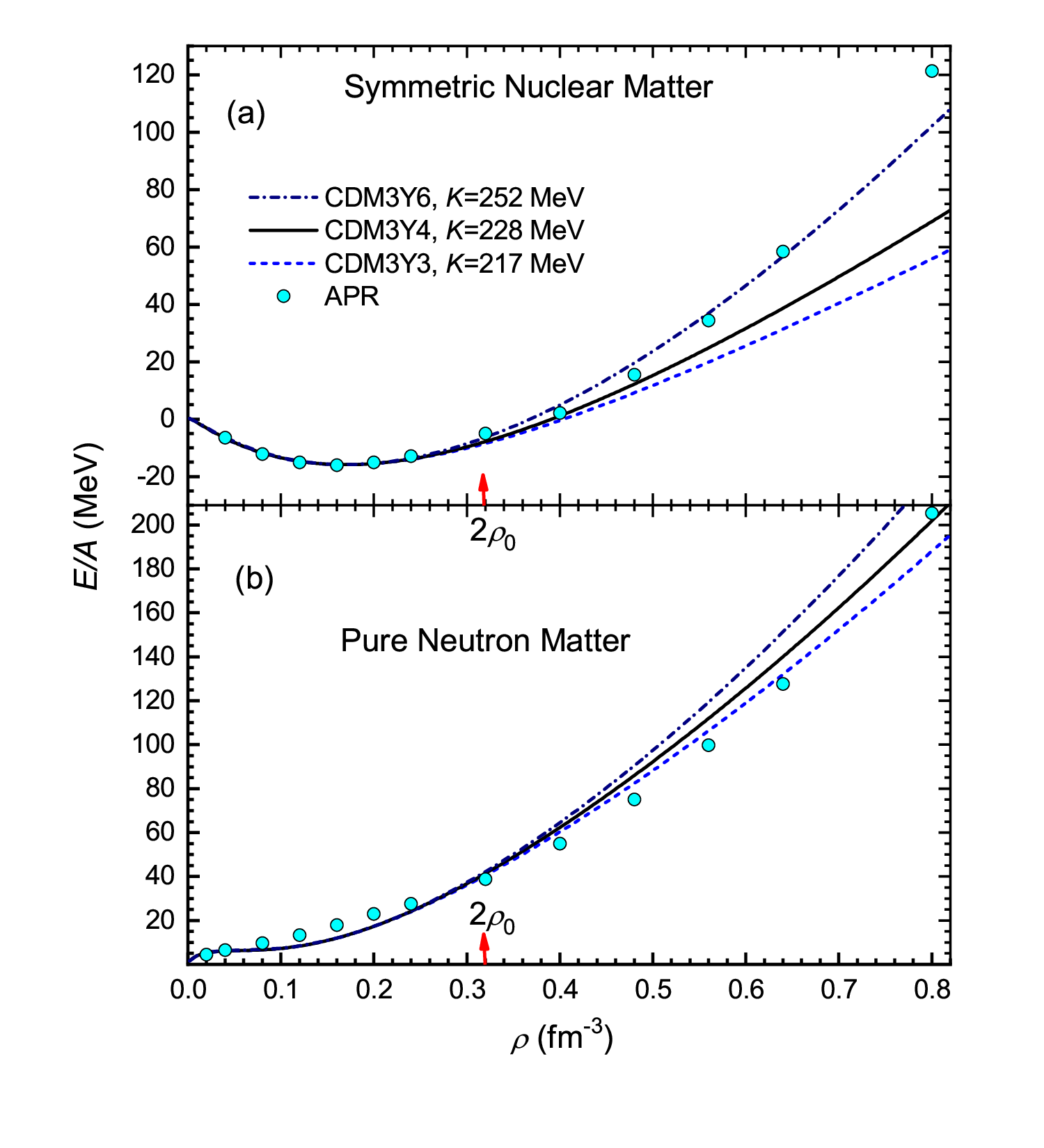}\vspace*{-1cm}
\caption{Energy per nucleon of the symmetric NM (a) and pure neutron matter (b) 
given by the HF calculation using the density- and isospin dependent CDM3Yn interaction. $K$ is 
the nuclear incompressibility obtained at the saturation density $\rho_0\approx 0.16$ fm$^{-3}$. 
The circles are results of the ab initio variational calculation by Akmal, Pandharipande and Ravenhall 
(APR) \cite{Ak98}. Arrow marks twice the saturation density, indicating the validity of the CDM3Yn 
interaction for the low-energy domain of nuclear EOS.} \label{f1}
\end{figure}

An effective density-dependent NN interaction is the essential input for the mean-field calculation 
of NM, and we recall briefly the CDM3Yn density dependent versions of the M3Y interaction 
used in our model. Originally, parameters of the CDM3Yn density dependence were parametrized 
\cite{Kho97} to reproduce the NM saturation properties on the HF level. These parameter sets were 
extended later to include a realistic isovector part  \cite{Loa11,Loa15,Kho07} as well as the rearrangement 
term of the single-particle potential \cite{Loa15}. The HF results for the energy per nucleon $E/A$ of NM 
obtained with the CDM3Yn interaction are compared in Fig.~\ref{f1} with results of the  \emph{ab initio} 
variational calculation using the Argonne V18 interaction \cite{Ak98}. One can see a nice agreement 
of the HF results with those of the ab initio calculation over a wide range of densities.  
We note that the functional CDM3Yn density dependence of  the original M3Y interaction \cite{An83} 
was parametrized \cite{Kho97} to reproduce on the HF level the saturation of symmetric NM at $\rho_0$, 
as well as to give a realistic (real) \AA OP within the double-folding model that properly describes 
the nuclear rainbow pattern observed in elastic \aA and light heavy-ion refractive scattering. 
The rainbow scattering data were proven to be sensitive to the real OP at small distances where
the dinuclear overlap density is reaching up to twice the saturation density $\rho_0$ \cite{Kho07r}, 
so that the use of the CDM3Yn interaction is well validated for the low-energy domain of nuclear EOS
(see Fig.~\ref{f1}).    

According to Landau theory for an infinite system of interacting fermions \cite{Mig67}, the single-particle 
energy is determined \cite{Loa15} as derivative of the energy per nucleon $\varepsilon\equiv E/A$ of NM  
with respect to the nucleon momentum distribution $n_\tau(k)$  
\begin{equation}
 E_{\tau}(\rho,k)=\frac{\partial \varepsilon}{\partial n_\tau(k)}=
\frac{\hbar^2 k^2}{2m_\tau}+U_{\tau}(\rho,k),\ {\rm where}\ \tau=n,p. \label{eq1}
\end{equation}
$E_{\tau}(\rho,k)$ is, thus, the change of the NM energy at density $\rho$ caused by the removal or 
addition of a nucleon with the momentum $k$. The single-particle potential $U_{\tau}(\rho,k)$ consists 
of both the HF and rearrangement terms 
\begin{equation}
 U_{\tau}(\rho,k)= U^{\rm (HF)}_{\tau}(\rho,k)+U^{\rm (RT)}_{\tau}(\rho,k), 
 \label{eq2}
\end{equation}
with the explicit expressions of $U_{\tau}(k)$ given in Ref.~\cite{Loa15}. At the Fermi momentum 
$(k\to k_{F\tau}),\ E_\tau(k_{F\tau})$ determined from Eqs.~(\ref{eq1})-(\ref{eq2}) is exactly 
the Fermi energy given by the Hugenholtz-van Hove (HvH) theorem \cite{HvH}, which is satisfied 
on the HF level only when the effective NN interaction is density independent, with the RT equal 
zero \cite{Loa15,Cze02}. In the mean-field calculation (\ref{eq1})-(\ref{eq2}), the RT originates 
naturally from the density dependence of the in-medium NN interaction that implicitly accounts 
for the higher-order NN correlations as well as three-body force \cite{Mah85,Zuo99,Zuo14}.

For the spin-saturated NM, the (spin-independent) direct (D) and exchange (EX) parts of the  
CDM3Yn interaction \cite{Kho97,Kho07} are used in the HF calculation of NM 
\begin{equation}
 v_{\rm D(EX)}(\rho,s)=F_0(\rho)v^{\rm D(EX)}_{00}(s) + 
 F_1(\rho)v^{\rm D(EX)}_{01}(s) (\bm{\tau}_1\cdot\bm{\tau}_2). \label{eq3}
\end{equation}  
The radial dependence of the isoscalar (IS) and isovector (IV) terms $v^{\rm D(EX)}_{00(01)}(s)$ 
is kept unchanged as that used for the original M3Y interaction \cite{An83}. The IS density dependence 
$F_0(\rho)$ was determined \cite{Kho97} to reproduce the saturation of NM on the HF level. 
The IV density dependence $F_1(\rho)$ was determined \cite{Loa15} based on the isospin 
dependence of nucleon OP given by the BHF calculation of NM by Jeukenne, Lejeune and Mahaux 
(JLM) \cite{Je77}, and and fine tuned by the folding model description of the charge exchange 
reaction to the isobar analog states in medium-mass nuclei \cite{Kho14}. 

Using the exact expression of the RT given by the HvH theorem, a compact method was suggested 
\cite{Loa15} to account for the RT on the HF level by adding a correction term to the CDM3Yn 
density dependence, $F_{0(1)}(\rho)\to F_{0(1)}(\rho)+\Delta F_{0(1)}(\rho)$, used in the 
HF-type calculation of the single-particle potential 
\begin{equation}
 U_{\tau}(\rho,k)=\sum_{k'\sigma'\tau'} 
 \langle \bm{k}\sigma\tau,\bm{k}'\sigma'\tau'|v_{\rm D}|
 \bm{k}\sigma\tau,\bm{k}'\sigma'\tau'\rangle +  
 \langle \bm{k}\sigma\tau,\bm{k}'\sigma'\tau'|v_{\rm EX}|
 \bm{k}'\sigma\tau,\bm{k}\sigma'\tau'\rangle,  \label{eq4}
\end{equation}
where $|\bm{k}\sigma\tau\rangle$ are the ordinary plane waves. Treating explicitly
the isospin dependence, the single-particle potential (\ref{eq4}) is expressed \cite{Loa15} 
in terms of the IS and IV parts as 
\begin{eqnarray}
U_\tau(\rho,k) &=& U_0^{\rm (HF)}(\rho,k)+U_0^{\rm (RT)}(\rho,k)
\pm [U_1^{\rm (HF)}(\rho,k)+U_1^{\rm (RT)}(\rho,k)] \nonumber \\
&=&[F_0(\rho)+\Delta F_0(\rho)]U^{\rm (M3Y)}_0(\rho,k)
\pm [F_1(\rho)\pm \Delta F_1(\rho)]U^{\rm (M3Y)}_1(\rho,k), \label{eq5}
\end{eqnarray}
where (-) sign pertains to proton and (+) sign to neutron. $U^{\rm (M3Y)}_0$ and $U^{\rm (M3Y)}_1$ 
are, respectively, the IS and IV parts of the single-particle potential given by the density independent 
M3Y interaction 
\begin{equation}
 U^{\rm (M3Y)}_{0(1)}(\rho,k)=\left[J^D_{0(1)}+\int\hat j_1(k_Fr)j_0(kr)
 v^{\rm EX}_{00(01)}(r)d^3r\right], \label{eq6} 
\end{equation}
\begin{equation}
{\rm where}\ J^{\rm D}_{0(1)}=\int v_{00(01)}^{\rm D}(r)d^3r, \
 \hat{j_1}(x)=3j_1(x)/x\ =\ 3(\sin x-x\cos x)/x^3. \nonumber 
\end{equation}
Because the original M3Y interaction is momentum independent,  the momentum dependence of the 
single-particle potential (\ref{eq5}) is due to the exchange term of $U^{\rm (M3Y)}_{0(1)}(\rho,k)$, 
with the nucleon momentum $k$ determined self-consistently as 
\begin{equation}
 k=\sqrt{\frac{2m_\tau}{\hbar^2}[E_\tau(\rho,k)-U_\tau(\rho,k)]}. \label{eq7}
\end{equation}

\begin{figure}[bht] \vspace*{-1cm}
\includegraphics[width=0.9\textwidth]{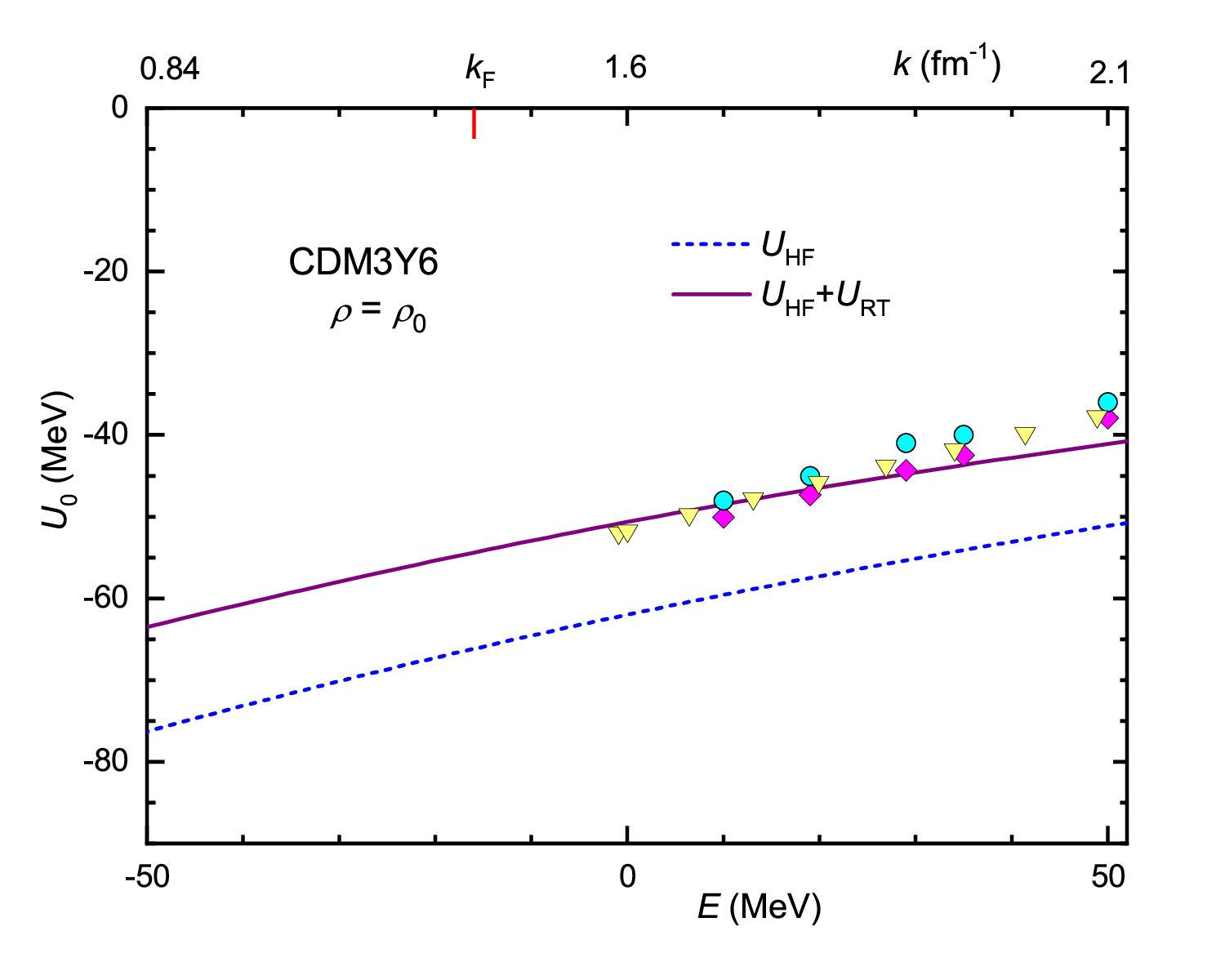}\vspace*{0cm}
 \caption{Single-particle potential in symmetric NM  (\ref{eq7}) determined at $\rho\approx \rho_0$ 
 with and without the RT using the CDM3Y6 interaction, in comparison with the empirical data 
 for nucleon OP taken from Refs.~\cite{BM69} (circles), \cite{Var91} (squares), and 
\cite{Hama} (triangles).} \label{f2}
\end{figure}
The extended HF approach (\ref{eq4})-(\ref{eq7}) provides a consistent description of both 
the single-particle potential for bound nucleons, $U_{\tau}(\rho,k)$ with $k<k_{F\tau}$, and nucleon 
optical potential, $U_{\tau}(\rho,k)$ with $k>k_{F\tau}$. Such an approach is the well-known 
continuous approximation for the single-particle potential \cite{Mah85,Mah91}, where nucleon OP 
in NM is determined as the mean-field potential felt by nucleon being scattered on NM at energy $E>0$. 
The momentum of scattered nucleon is determined by the same relation (\ref{eq7}) but with $E_\tau$ 
replaced by the nucleon incident energy $E$. In this way, the energy- and momentum dependence 
of nucleon OP are treated on the same footing as illustrated in Fig.~\ref{f2}.  Therefore, an important 
constraint for the present approach is that at $E>0$ the energy dependence of the potential (\ref{eq5}) 
should agree reasonably with the observed energy dependence of nucleon OP.  The single-particle 
potential (\ref{eq5}) evaluated for symmetric NM at the saturation density $\rho_0$ using the 
CDM3Y6 interaction is compared with the empirical data \cite{BM69,Var91,Hama} for nucleon OP 
at $E > 0$ in Fig.~\ref{f2}. One can see that the inclusion of the RT significantly improves 
the agreement with the empirical data at energies $E<50$ MeV. Moreover, the momentum dependence 
of nucleon OP in this low-energy region is found nearly linear and due mainly to the Pauli exchange term 
in Eq.~(\ref{eq4}).  

It should be noted here that the Pauli exchange term in Eq.~(\ref{eq4}) gives a good agreement 
of the energy (or momentum) dependence of nucleon OP with the empirical data at low energies
only. At higher energies ($E>50$ MeV or $k>2$ fm$^{-1}$) the agreement becomes worse, 
and nucleon OP (\ref{eq5}) is more attractive compared to the empirical data (see, e.g., Fig.~3 in 
Ref.~\cite{Kho16}). Such a behavior of nucleon OP given by the HF calculation is well expected 
in light of the effective G-matrix interaction derived from the solution of  Brueckner–Bethe–Goldstone 
equation \cite{Ken01}, where both the \emph{direct} and \emph{exchange} parts of the G matrix 
interaction are energy dependent. This is also the reason why a slight linear energy dependence was added 
to the CDM3Y6 interaction, in terms of $g(E)$ factor \cite{Kho97}, for the double-folding calculation 
of \AA OP at medium energies. Consequently, the nucleon effective mass $m^*$  at high momenta 
originates not only from the Pauli nonlocality, but also from the higher-order NN correlations in 
nuclear medium. The present work is focused, however, on the determination of  $m^*$  at low 
energies ($E<50$ MeV) and corresponding momenta $k<2$ fm$^{-1}$ from the momentum 
dependence of nucleon OP, which is due entirely to the Pauli nonlocality as shown above.   

\section{Folding model of nucleon optical potential}
\subsection{Nonlocal folded nucleon OP}
\label{sec2} 
The folding model of nucleon OP \cite{Ken01,Kho02,Minomo} is known to generate
the first-order term of the microscopic nucleon OP defined in Feshbach's formalism 
of nuclear reactions \cite{Fe92}. The success of the folding approach in the OM description 
of elastic \nA scattering at low and medium energies confirms that the folded nucleon OP
is the dominant part of the microscopic nucleon OP. In a consistent mean-field consideration, 
the central OP for elastic nucleon scattering on a target nucleus $A$ is evaluated using the same 
Eq.~(\ref{eq4}), but with plane waves $|\bm{k}'\sigma'\tau'\rangle$ replaced by the single-particle 
wave functions $|j\rangle$ of target nucleons 
\begin{equation} 
  U(k)=\sum_{j\in A}[\langle \bm{k},j|v_{\rm D}|\bm{k},j\rangle 
	+\langle \bm{k},j|v_{\rm EX}|j,\bm{k}\rangle]. \label{eq8}
\end{equation}
The antisymmetrization of the \nA system is done in the HF manner, taking into account 
explicitly the nucleon knock-on exchange. As a result, the exchange term of the folded nucleon OP
(\ref{eq8}) becomes \emph{nonlocal} in coordinate space \cite{Ken01}, and the OM equation 
for elastic nucleon scattering at energy $E$ becomes an integro-differential equation 
\begin{eqnarray}
 && \left[-\frac{\hbar^2}{2\mu}\nabla^2+U_{\rm D}(R)+V_{\rm C}(R)
 +V_{\rm s.o.}(R)(\bm{L}\cdot\bm{\sigma})\right]\Psi(\bm{R}) \nonumber\\ 
 &&\hskip 2.6cm +\int K(\rho,\bm{R},\bm{r})
 \Psi(\bm{r})d^3r= E~\Psi(\bm{R}), \label{eq9}
\end{eqnarray}
where $V_{\rm s.o.}(R)$ is the spin-orbit potential and $V_{\rm C}(R)$ is the Coulomb potential 
used for elastic proton scattering only. The scattering wave function $\Psi(\bm{R})$ is obtained from 
the solution of the OM equation (\ref{eq9}) at each \nA distance $R$. Based on the CDM3Yn interaction
(\ref{eq3}) with the RT included, the mean-field part of nucleon OP consists of the local direct potential 
$U_{\rm D}(R)$ and the exchange integral involving a nonlocal density-dependent kernel 
$K(\rho,\bm{R},\bm{r})$. Making explicit the isospin degrees of freedom, the mean-field part of nucleon
OP can be expressed (in the Lane manner) in terms of the IS and IV components as   
\begin{eqnarray}
 && U_{\rm D}(R)=U^{\rm D}_{\rm IS}(R)\pm U^{\rm D}_{\rm IV}(R), \nonumber\\ 
 && K(\rho,\bm{R},\bm{r})=K_{\rm IS}(\rho,\bm{R},\bm{r})
 \pm K_{\rm IV}(\rho,\bm{R},\bm{r}), \label{eq10}  
\end{eqnarray}
where (-) sign pertains to proton OP and (+) sign to neutron OP. The IS and IV  parts of the direct 
folded potential and nonlocal exchange kernel (\ref{eq10}) are given through the IS and IV nucleon density 
matrices, respectively, as  
\begin{eqnarray}
&& U^{\rm D}_{\rm IS(IV)}(R)=\int\big[\rho_n(\bm{r})\pm\rho_p(\bm{r})\big]
v^{\rm D}_{00(01)}(\rho,s)d^3r,  \nonumber \\  
&& K_{\rm IS(IV)}(\rho,\bm{R},\bm{r})=\big[\rho_n(\bm{R},\bm{r})\pm\rho_p(\bm{R},\bm{r})\big]
v^{\rm EX}_{00(01)}(\rho,s), \label{eq11}
\end{eqnarray}
where $s=|\bm R-\bm r|$. The nucleon density matrix is determined from the single-particle wave functions 
$\varphi^{(\tau)}_j$ of target nucleons as
\begin{equation}
 \rho_\tau(\bm{R},\bm{r})=\sum_{j\in A}\varphi^{(\tau)*}_j(\bm{R})
 \varphi^{(\tau)}_j(\bm{r}),\ {\rm with}\  \rho_\tau(\bm r)\equiv \rho_\tau(\bm{r},\bm{r})\ {\rm and}  
 \ j\equiv nlj. \label{eq12}
\end{equation}
The direct potential $U_{\rm D}(R)$ is readily obtained by folding the nucleon densities with 
the direct part $v^{\rm D}_{00(01)}(\rho,s)$ of the density dependent CDM3Yn interaction (\ref{eq3}), 
including the contribution of the RT, as
\begin{equation} \label{eq13}
U^{\rm D}_{\rm IS(IV)}(R)=\int\big[\rho_n(\bm{r})\pm\rho_p(\bm{r})\big]
\big[F_{0(1)}(\rho(\bm r))\pm\Delta F_{0(1)}(\rho(\bm r))\big]v^{\rm D}_{00(01)}(s)d^3r,
\end{equation}
where the $(\pm)$ signs are used in the same way as in Eqs.~(\ref{eq5}) and (\ref{eq10}), and 
the contribution of the RT to the IV part of the direct potential $U^{\rm D}_{\rm IV}$ via 
$\Delta  F_1(\rho)$ is the same for both the proton and neutron OP. The exact treatment of the nonlocal 
exchange term is cumbersome and involves the explicit angular-momentum dependence of the exchange 
kernel. Using the multipole decomposition of the radial Yukawa function of the exchange part 
of the CDM3Yn interaction (\ref{eq3}) 
\begin{equation}  
 v^{\rm EX}_{00(01)}(s)=\sum_{\lambda\mu}\frac{4\pi}{2\lambda+1}
 X^{(\lambda)}_{00(01)}(R,r)Y^*_{\lambda\mu}(\hat{\bm R})
 Y_{\lambda\mu}(\hat{\bm r}), \label{eq14} 
\end{equation}
we obtain, after integrating out the angular dependence, the radial equation for each partial wave 
\begin{eqnarray} 
& &-\frac{\hbar^2}{2\mu}\left[\frac{d^2}{dR^2}-\frac{L(L+1)}{R^2}\right]
\chi_{LJ}(R)+\big[U_{\rm D}(R)+V_{\rm C}(R) \nonumber\\
& &+A_{LJ}V_{\rm s.o.}(R)\big]\chi_{LJ}(R) +\int_0^\infty K_{LJ}(\rho,R,r)
\chi_{LJ}(r)dr=E~\chi_{LJ}(R), \label{eq15}
\end{eqnarray}
where $\chi_{LJ}(R)/R$ is the radial part of nucleon scattering wave function $\Psi(\bm R)$,
$A_{LJ}$ is the s.o. coupling coefficient determined as $A_{LJ}=L$ if $J=L+1/2$, and 
$A_{LJ}=-L-1$ if $J=L-1/2$. The nonlocal density-dependent exchange kernel is then obtained
at each partial wave as  
\begin{eqnarray}
 K_{LJ}(\rho,R,r)&=&\big[K^{\rm IS}_{LJ}(\rho,R,r)\pm K^{\rm IV}_{LJ}(\rho,R,r)\big], \label{eq16} \\
 K^{\rm IS}_{LJ}(\rho,R,r)&=&\big[F_0(\rho(r))+\Delta F_0(\rho(r))\big]
 \sum_{nlj,\lambda}\left[u^{(n)}_{nlj}(R) u^{(n)}_{nlj}(r) + u^{(p)}_{nlj}(R) u^{(p)}_{nlj}(r)\right] 
 \nonumber\\  & & \times (2j+1) X^{(\lambda)}_{00}(R,r)
 \left(\begin{array}{l}L\\0\end{array}
 \begin{array}{l}l\\0\end{array}
 \begin{array}{l}\lambda\\0\end{array}\right)^2, \label{eq17} \\ 
 K^{\rm IV}_{LJ}(\rho,R,r)&=&\big[F_1(\rho(r))\pm\Delta F_1(\rho(r))\big]
 \sum_{nlj,\lambda}\left[u^{(n)}_{nlj}(R)u^{(n)}_{nlj}(r)-u^{(p)}_{nlj}(R) u^{(p)}_{nlj}(r)\right] 
\nonumber\\  & &\ \times (2j+1) X^{(\lambda)}_{01}(R,r)
 \left(\begin{array}{l}L\\0\end{array}
 \begin{array}{l}l\\0\end{array}
 \begin{array}{l}\lambda\\0\end{array}\right)^2. \label{eq18}  
\end{eqnarray}
Here $u^{(\tau)}_{nlj}(r)/r$ is the radial part of the single-particle wave function 
$\varphi^{(\tau)}_{nlj}(\bm{r})$ of target nucleon. The (-) sign is used with proton OP and (+) sign 
with neutron OP, and the contribution of the RT to the IV part of the exchange kernel is also the same 
for both proton and neutron OP as found for the IV part of the direct potential (\ref{eq13}). The explicit 
representation of nucleon OP in terms of the IS and IV parts should be helpful in revealing the contribution 
of valence neutrons to the total OP. Furthermore, the charge exchange form factor (FF) of the $(p,n)$ 
reaction to the isobar analog state (IAS) is determined, in the Lane isospin coupling scheme, entirely 
by the IV part of nucleon OP \cite{Kho14,Loc14}. Therefore, the present nonlocal folding model can also 
be used to calculate the nonlocal charge exchange FF of the Fermi transition to IAS in the folding model 
study of $(p,n)$IAS reaction.     

\subsection{Local approximation for the folded nucleon OP}
\label{sec3}
Although the nucleon OP is well established to be nonlocal in the coordinate space due to the Pauli blocking 
(as discussed above) and nonelastic channel coupling, the nucleon OP of some local form is widely used 
in numerous OM analyses of elastic \nA scattering. The obtained elastic scattering wave function (dubbed 
as the distorted wave) is then used as a key input in different DWBA or coupled-channel calculations 
of direct nuclear reactions. We briefly discuss here the validity of the local folding model \cite{Kho02,Kho07} 
of nucleon OP, which has been successfully used to study elastic nucleon scattering. 

Applying a local WKB approximation \cite{Sin75,Bri77} for the change in the scattering wave function 
in the OM equation (\ref{eq9}) induced by the exchange of spatial coordinates of scattered nucleon 
and target nucleon, we obtain 
\begin{equation}
\Psi(\bm{r})=\Psi(\bm{R+s})\simeq \Psi(\bm{R})
 \exp\big(i{\bm k}(E,\bm R)\cdot{\bm s}\big), \label{eq19} 
\end{equation}
where the local momentum $k(E,R)$ of scattered nucleon is energy- and radial dependent. Inserting
Eq.~(\ref{eq19}) into Eq.~(\ref{eq9}), the nonlocal exchange integral becomes \emph{localized} and
\emph{energy (momentum) dependent}. The local exchange term of the folded nucleon OP can then be 
evaluated separately using the nonlocal nucleon density matrix     
\begin{eqnarray}
U_{\rm EX}(E,R,k)&=&U^{\rm EX}_{\rm IS}(E,R,k)\pm U^{\rm EX}_{\rm IV}(E,R,k),  \nonumber\\ 
U^{\rm EX}_{\rm IS(IV)}(E,R,k)&=& \int\big[\rho_n(\bm{R},\bm{r})\pm\rho_p(\bm{R},\bm{r})\big] 
 j_0\big(k(E,R)s\big) v^{\rm EX}_{00(01)}(\rho,s)d^3r  \nonumber\\ 
 &=& \int\big[\rho_n(\bm{R},\bm{r})\pm\rho_p(\bm{R},\bm{r})\big]
\big[F_{0(1)}(\rho(\bm r))\pm\Delta F_{0(1)}(\rho(\bm r))\big] \nonumber \\ 
 & &\ \  \times j_0\big(k(E,R)s\big) v^{\rm EX}_{00(01)}(s)d^3r, \label{eq20}
\end{eqnarray}
where the $(\pm)$ signs are used in the same way as in Eq.~(\ref{eq10}). The local momentum 
of scattered nucleon is determined self-consistently from the real folded nucleon OP as
\begin{equation}
 k^2(E,R)=\frac{2\mu}{\hbar^2}\left\{E-{\rm Re}\left[U_{\rm D}(R)+U_{\rm EX}(E,R,k)\right]-
V_{\rm C}(R)\right\}.\label{eq21}
\end{equation}  
Thus, the momentum- and energy dependences of the local folded nucleon OP of finite nuclei are
directly interrelated and also determined on the same footing as in the NM limit (see Fig.~\ref{f2}). 
The method to evaluate the direct (\ref{eq13}) and exchange (\ref{eq20}) folded potentials is given 
in more details in, e.g., Ref.~\cite{Kho02}. Using a realistic local approximation for the nonlocal 
density matrix in Eq.~(\ref{eq20}), the nuclear density $\rho(\bm r)$ obtained in any structure 
model or directly deduced from electron scattering data can be used in the folding model calculation 
of the local nucleon OP. 

\subsection{Elastic nucleon scattering on $^{40,48}$Ca, $^{90}$Zr, and  $^{208}$Pb targets}
\label{sec4} 
To validate the use of the local folded nucleon OP to deduce the nucleon effective mass, 
we show here results of the folding model analysis of low-energy elastic nucleon scattering 
using both the nonlocal and local folded nucleon OP. Namely, the nonlocal and local folding models 
of nucleon OP discussed above are used to calculate nucleon OP for the OM study of elastic neutron 
and proton scattering on $^{40,48}$Ca, $^{90}$Zr, and $^{208}$Pb targets. 
Because the nonlocal folding calculation (\ref{eq11})-(\ref{eq12}) requires the explicit single-particle 
wave functions of target nucleons, we have used the single-particle wave functions given by the HF 
calculation of finite nuclei based upon a complete basis of spherical Bessel functions \cite{Tha11} 
and finite-range D1S Gogny interaction \cite{Ber91}.     

\begin{figure}[bht] \vspace*{-1cm}
\includegraphics[width=1.1\textwidth]{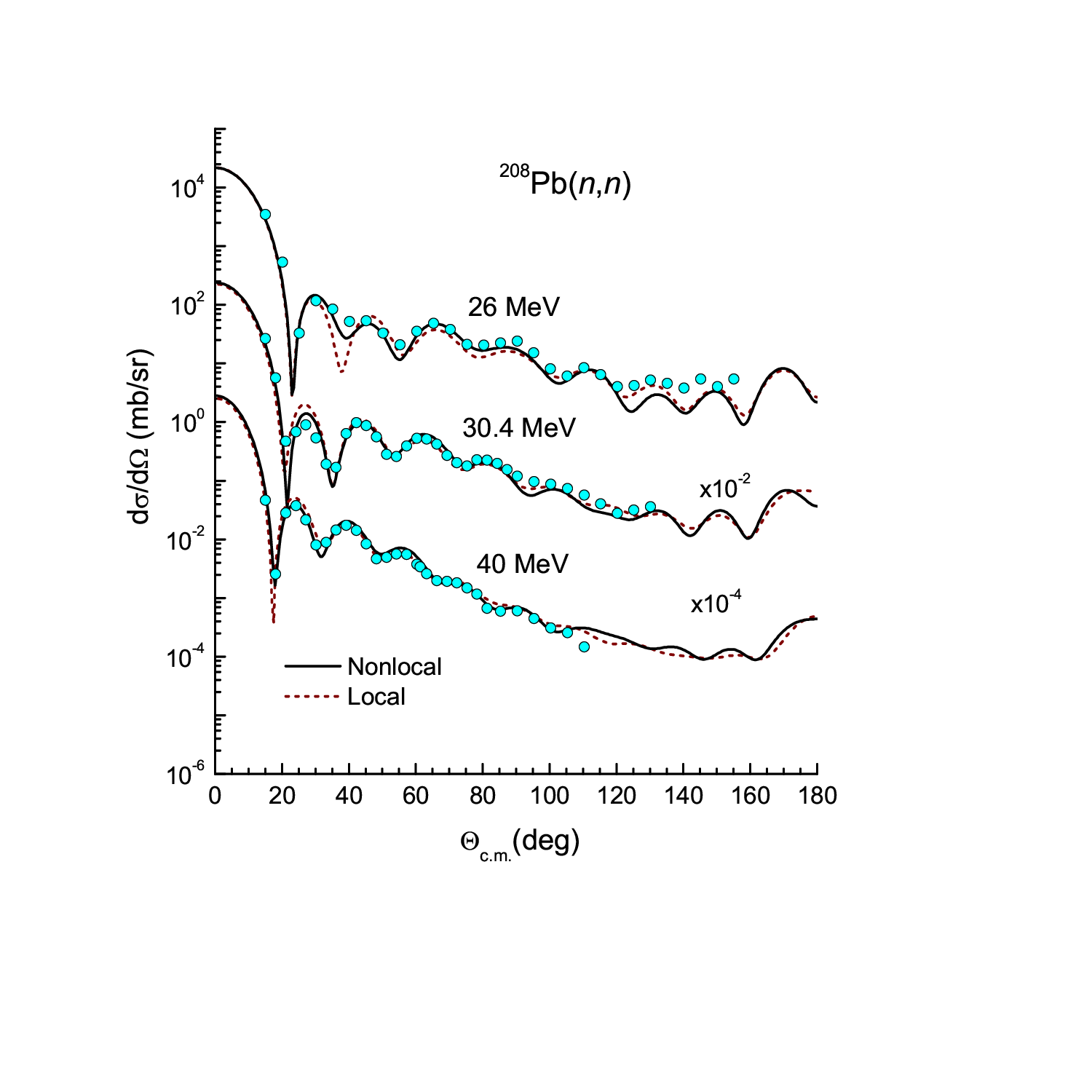}\vspace*{-4cm}
 \caption{OM description of elastic \nPb scattering data measured at 26, 30, and 40 MeV 
\cite{nPb1,nPb2,nPb3} given by the nonlocal and local folded neutron OP's obtained with the 
complex CDM3Y6 interaction \cite{Loa20}.} \label{f3}
\end{figure}
To obtain a \emph{complex} folded nucleon OP, it is necessary to have a realistic complex parametrization 
of the density dependent CDM3Yn interaction. For this purpose, the imaginary density dependence 
of the CDM3Y6 interaction was determined using the same density dependent functional $F_{0(1)}(\rho)$ 
as that used for the real interaction (\ref{eq3}), with the parameters adjusted to reproduce, at each energy, 
the imaginary nucleon OP in NM given by the JLM parametrization of the BHF results \cite{Je77} 
on the HF level \cite{Kho07}. The complex nonlocal and local folded OP's were then used as input for 
the OM calculation of elastic nucleon scattering using the calculable $R$-matrix method \cite{desco,desco2}. 
In the present OM calculation, both the nonlocal and local folded OP's are supplemented by the local 
Coulomb and spin-orbit potentials taken from the global systematics CH89 of nucleon OP \cite{Var91}. 

\begin{figure}[bht] \vspace*{-1cm}
\includegraphics[width=0.8\textwidth]{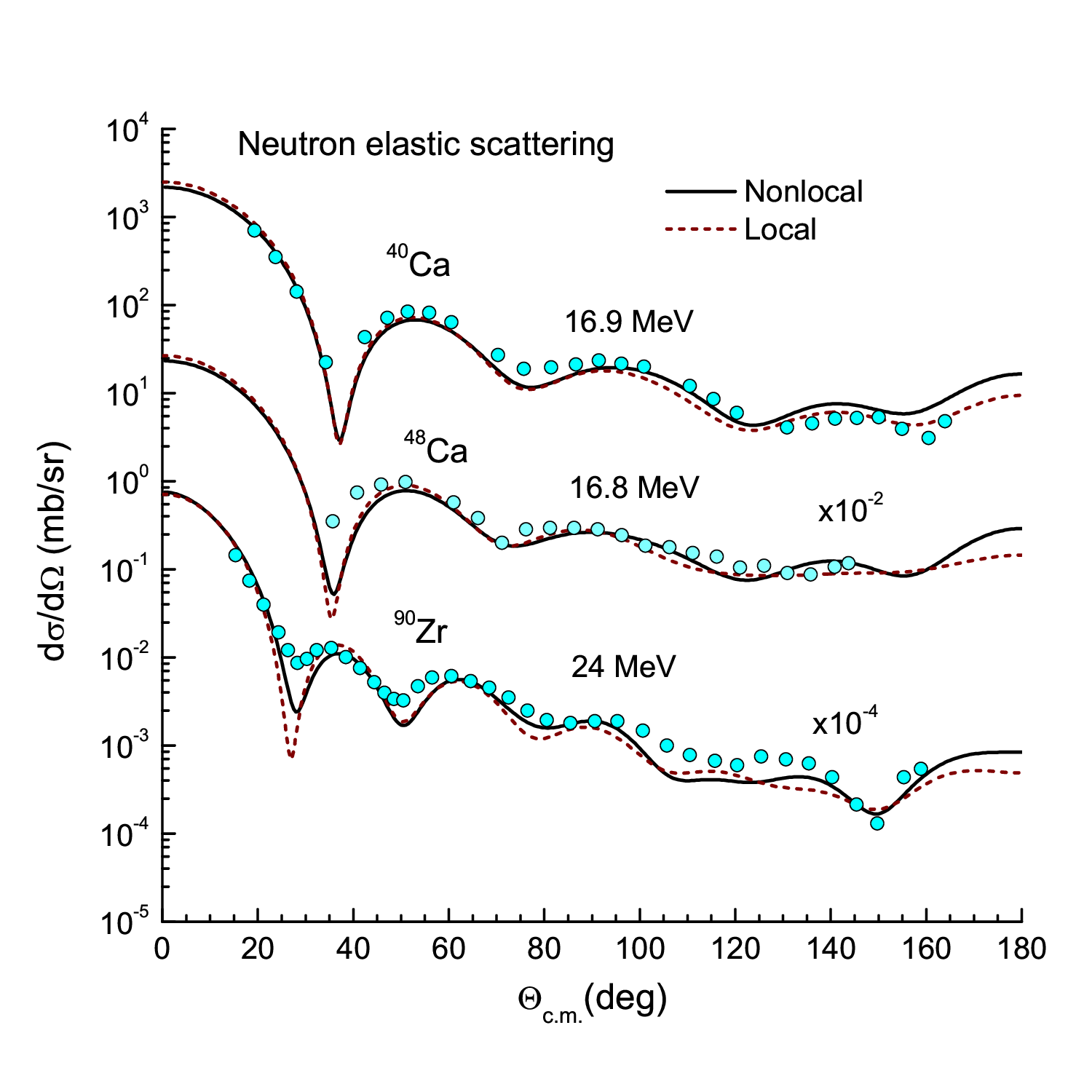}\vspace*{-1cm}
 \caption{The same as Fig.~\ref{f3} but for elastic neutron scattering data measured at 17 and 24 MeV
 \cite{ndat1,ndat2,ndat3} for $^{40,48}$Ca and $^{90}$Zr targets, respectively.} \label{f4}
\end{figure}
\begin{figure}[bht] \vspace*{-1cm}
\includegraphics[width=1.1\textwidth]{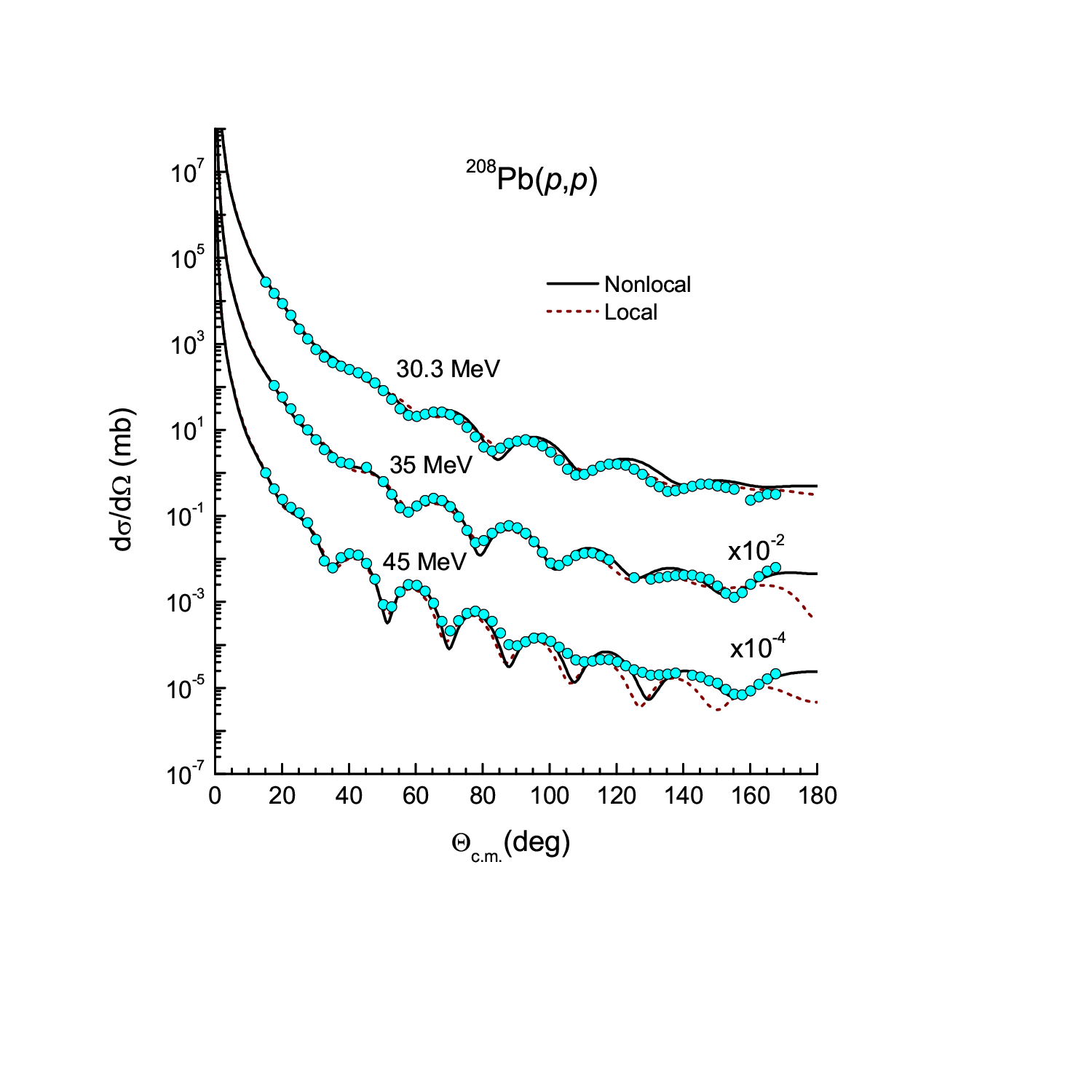}\vspace*{-4cm}
 \caption{The same as Fig.~\ref{f3} but for elastic \pPb scattering data measured at 30, 35, and 
45 MeV \cite{pPb1,pPb2}.} \label{f5}
\end{figure}
The reliability of the folded OP is best to be probed in the OM analysis of elastic neutron scattering 
from a heavy target at low energies, where the Coulomb interaction is absent and the mean-field 
dynamics is well established. For this purpose, elastic \nPb scattering data measured accurately 
at energies of 26, 30.4, and 40 MeV  \cite{nPb1,nPb2,nPb3} turned out to be a very good test 
ground.  Given parameters of the real CDM3Yn interaction fine-tuned by the HF description of NM 
shown in Fig.~\ref{f1}, no renormalization of the \emph{real} part of both the nonlocal and local 
folded OP was allowed in the present OM study to test its proximity to the real nucleon OP. The 
imaginary folded OP based on the JLM parametrization delivers a good OM description of elastic 
proton scattering, but it gives a stronger absorption of neutron OP, and an overall renormalization 
of the \emph{imaginary} part of both nonlocal and local folded neutron OP by a factor $\sim 0.8$ was 
found necessary for a good OM description of the considered neutron scattering data. From the OM 
results obtained for elastic \nPb scattering shown in Fig.~\ref{f3} one can see that both the nonlocal 
and local folded OP deliver the same good OM description of data over the whole angular 
range, which validates the local approximation (\ref{eq20})-(\ref{eq21}) for the exchange term of the 
folded neutron OP. We note that, in the absence of the Coulomb interaction, the (diffractive) 
oscillation of elastic neutron scattering cross section at small angles can be properly reproduced 
only with the inclusion of the RT into the folding calculation \cite{Loa20}.  
A good accuracy of the local approximation for the folded OP can also be seen in the OM 
results obtained for elastic neutron scattering on the medium-mass $^{40,48}$Ca and $^{90}$Zr 
targets shown in Fig.~\ref{f4}.    

\begin{figure}[bht] \vspace*{-1cm}
\includegraphics[width=1.1\textwidth]{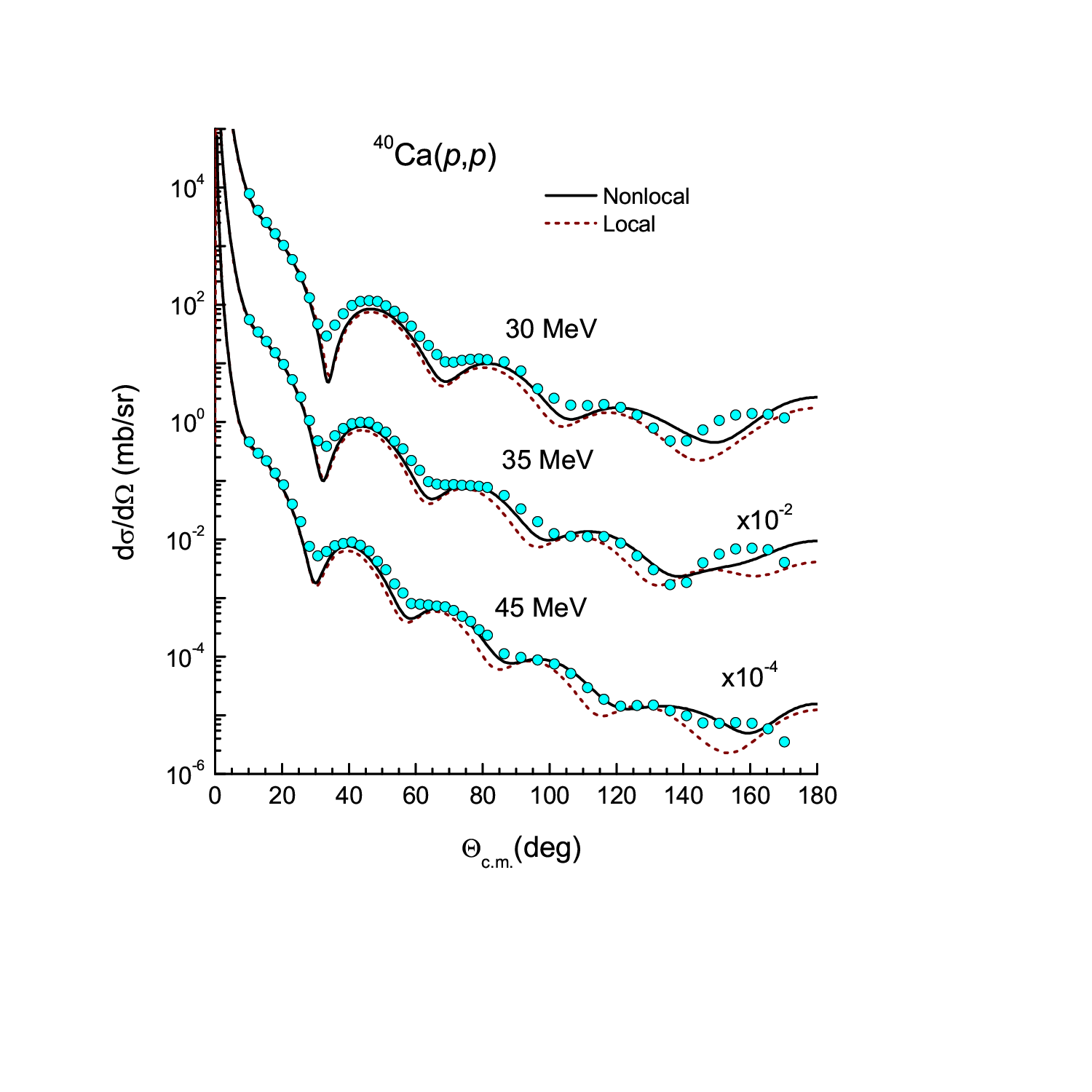}\vspace*{-4cm}
 \caption{The same as Fig.~\ref{f3} but for elastic \pCa scattering data measured at 30, 35, and 
 45 MeV \cite{nCa}.} \label{f6}
\end{figure}
The elastic \pPb scattering data measured at 30.4, 35, and 45 MeV \cite{pPb1,pPb2} are compared 
in Fig.~\ref{f5} with the OM results given by the folded proton OP. Both the nonlocal and local folded 
OP were found to give the same good OM description of elastic \pPb data at the forward and medium 
angles, while the data points at backward angles are better reproduced by the nonlocal folded OP, 
especially, at 45 MeV. We note that the OM results shown in Figs.~\ref{f5} and \ref{f6} were obtained 
without renormalizing the (complex) strength of the folded proton OP.  Like the OM result obtained for 
elastic \pPb scattering at $E=45$ MeV, one can also see a more pronounced difference given by the 
nonlocal folded proton OP at medium and large scattering angles from the results obtained for elastic 
\pCa scattering at energies $E\gtrsim 30$ MeV shown in Fig.~\ref{f6}, which indicates the need of 
taking into account exactly the Pauli nonlocality of the folded proton OP at these medium energies.  

\section{Nucleon effective mass}
\label{sec5}
As discussed in Sec.~\ref{sec1}, the momentum dependence of the single-particle potential in NM 
at low momenta is determined mainly by the exchange term which results on a \emph{nonlocal} 
single-nucleon potential in the coordinate space. At positive energy ($E>0$) the single-nucleon 
potential (\ref{eq4})-(\ref{eq5}) can be treated as nucleon OP in NM, with the associated nucleon 
effective mass $m^*_\tau$ determined \cite{Mah85,Hog83} as
\begin{equation}
\frac{m^*_\tau(\rho,\delta,k)}{m}=\Biggl[1+\frac{m}{\hbar^2k}
\frac{\partial U_\tau(\rho,\delta,k)}{\partial k}\Biggr]^{-1}, \ \mbox{with}\ \tau=n,p \label{eq22}
\end{equation}
where $m$ is the free nucleon mass. In general, the nucleon effective mass arises from both the 
momentum- and energy dependence of the single-particle potential \cite{Mah85}, known as the 
$k$- and $E$-effective masses, which characterize the spacetime nonlocality. Within the time 
independent HF formalism \cite{Loa15}, the total nucleon effective mass is determined by the same 
relation (\ref{eq22}), and it is associated with the spatial nonlocality of the nucleon mean-field  
potential. At the Fermi momentum ($k\to k_{F\tau}$) , the nucleon effective mass (\ref{eq22}) is 
obtained naturally from the Fermi energy (\ref{eq1}) by the Hugenholtz-van Hove theorem \cite{HvH}. 
The knowledge about $m^*_\tau$ at the saturation density $\rho_0$ and different neutron-proton 
asymmetries $\delta=(\rho_n-\rho_p)/\rho$ is essential for the determination of the nuclear 
symmetry energy $E_{\rm sym}$ and its slope parameter $L$, the two key ingredients of the EOS 
of neutron rich NM \cite{Li18,Hor14}. 
\begin{figure}[bht] \vspace*{-0.5cm}
\includegraphics[width=1.1\textwidth]{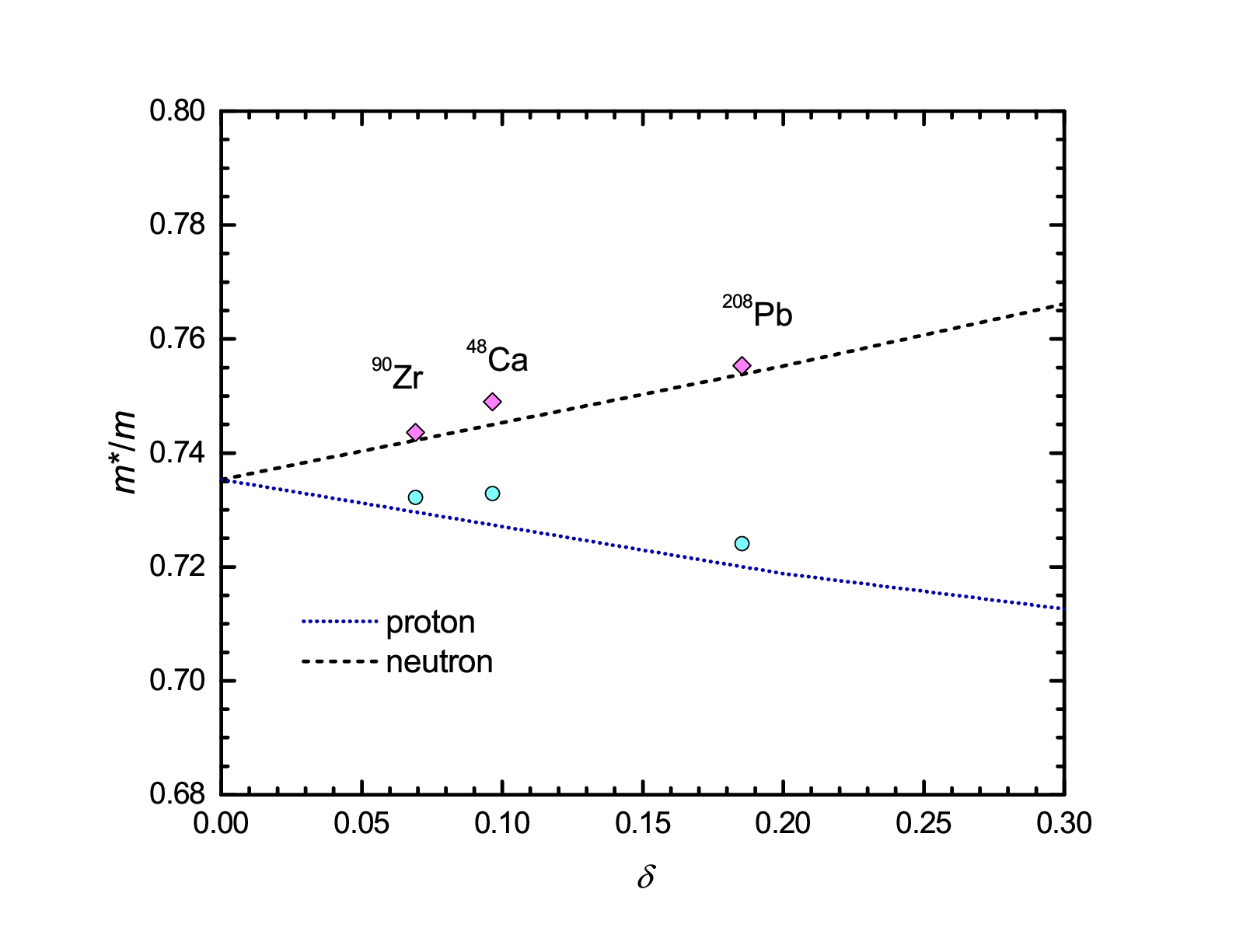}\vspace*{-1cm}
\caption{Neutron- and proton effective masses (\ref{eq22}) given by the extended HF calculation of 
asymmetric NM using the CDM3Y6 interaction (dashed and dotted lines, respectively) at 
$\rho=\rho_0,\ k= k_{F\tau}$, and different neutron-proton asymmetries $\delta$. The symbols are those 
obtained from the real folded nucleon OP for finite nuclei at $\rho\approx\rho_0$, $k\gtrsim k_{F\tau}$ 
(see Table~\ref{t1}).} 
\label{f7}
\end{figure}

The neutron and proton effective masses (\ref{eq22}) obtained from the single-particle potential 
(\ref{eq4})-(\ref{eq5}) in asymmetric NM at the saturation density $\rho_0$ with $k\to k_{F\tau}$ 
are shown in Fig.~\ref{f7} as dashed and dotted lines, respectively. 
Although the $m^*_\tau$ values are still poorly known at high NM densities and/or large $\delta$ 
values, the empirical nucleon effective mass in symmetric NM ($\delta=0$) at the saturation density 
$\rho_0$ is well established to be around $m^*/m\approx 0.73$ \cite{Hog83}. Our extended HF calculation 
of NM using the CDM3Yn interaction gives $m^*/m\approx 0.737$ at $\rho\approx\rho_0$ and $\delta=0$ 
(see Fig.~\ref{f9}), in a good agreement with the empirical data. The isospin dependence of the nucleon 
effective mass is governed by the neutron-proton effective mass splitting 
\begin{equation}
 m^*_{n-p}=(m^*_n-m^*_p)/m, \label{eq23}
\end{equation}
which is associated directly with the nuclear symmetry energy and its slope parameter. The knowledge 
of $m^*_{n-p}$ is also of importance for the determination of the neutron/proton ratio during stellar 
evolution or cooling of protoneutron star \cite{Li13,Li15,Li18}. Depending on the isospin dependence 
of the in-medium NN interaction, one obtains very different results for the neutron-proton effective mass 
splitting. A survey of different mean-field studies in Ref.~\cite{Li18} shows that $m^*_{n-p}$ values 
depend linearly on the neutron-proton asymmetry parameter $\delta$, and are ranging widely from 
negative- to positive values, up to $m^*_{n-p}\approx 0.637~\delta$. An analysis of the empirical 
constraints for the density dependence of nuclear symmetry energy  \cite{Li13} from both the nuclear 
physics experiments and astrophysical observations has led to the empirical constraint 
$m^*_{n-p}(\rho_0,\delta)\approx (0.27\pm 0.25)\delta$. Our extended HF calculation 
of asymmetric NM using the CDM3Y6 interaction gives $m^*_{n-p}(\rho_0,\delta)\approx 
(0.20\pm 0.02)\delta$ at the Fermi momentum, which is well inside the empirical boundary. 

For many years, the nucleon effective mass in finite nuclei has been a focus of different nonrelativistic 
and relativistic nuclear structure studies (see, e.g., Refs.~\cite{Giai83,Lit06,Zal10}) which provided 
accurate estimates of the $m^*_\tau$ values for the single-particle states of bound nucleons 
(with $k\lesssim k_{F\tau}$). Although the neutron- and proton effective masses, and the corresponding 
single-particle energies and spectroscopic factors are well described by the structure studies, it remains 
difficult to deduce therefrom an explicit isospin dependence of $m^*_\tau$ or the neutron-proton effective 
mass splitting (\ref{eq23}).     
The method often used so far for this purpose is to deduce the isospin dependence of the nucleon effective 
mass from the isospin dependence of the phenomenological nucleon OP given by the best OM fit to 
elastic nucleon scattering data at different energies \cite{XLi15}. In the present work, we aim to determine 
the nucleon effective mass from the semi-microscopic nucleon OP predicted by the local folding model 
presented in Sec.~\ref{sec3}. Given a good OM description of elastic nucleon scattering at low energies 
by the local folded nucleon OP shown Sec.~\ref{sec4}, it is reasonable to determine the effective mass 
of nucleon scattered by the mean field of target from the momentum dependence of the real folded OP as 
\begin{eqnarray}
\frac{m^*_\tau(E,R,k)}{m}&=&\left\{1+\frac{m}{\hbar^2k}
\left[\frac{\partial V_\tau(E,R,k)}{\partial k}\right]\right\}^{-1},  \label{eq24}\\
\mbox{where}\ V_\tau(E,R,k)&=&{\rm Re}\left[U^{(\tau)}_{\rm D}(R)+
 U^{(\tau)}_{\rm EX}(E,R,k)\right].  \label{eq25}
\end{eqnarray}
Here $U^{(\tau)}_{\rm D(EX)}$ are the direct and exchange parts of the local neutron (or proton) folded OP 
determined from Eqs.~(\ref{eq13}) and (\ref{eq20}), respectively. The energy or momentum dependence 
of the folded nucleon OP (\ref{eq25}) is embedded in the exchange term only, which was shown in 
Sec.~\ref{sec3} to be resulted from the Pauli nonlocality of the folded nucleon OP. The in-medium momentum 
$k$ of scattered nucleon is determined self-consistently from the real folded nucleon OP by Eq.~(\ref{eq21}). 
\begin{figure}\vspace*{-1cm}
\includegraphics[width=\textwidth]{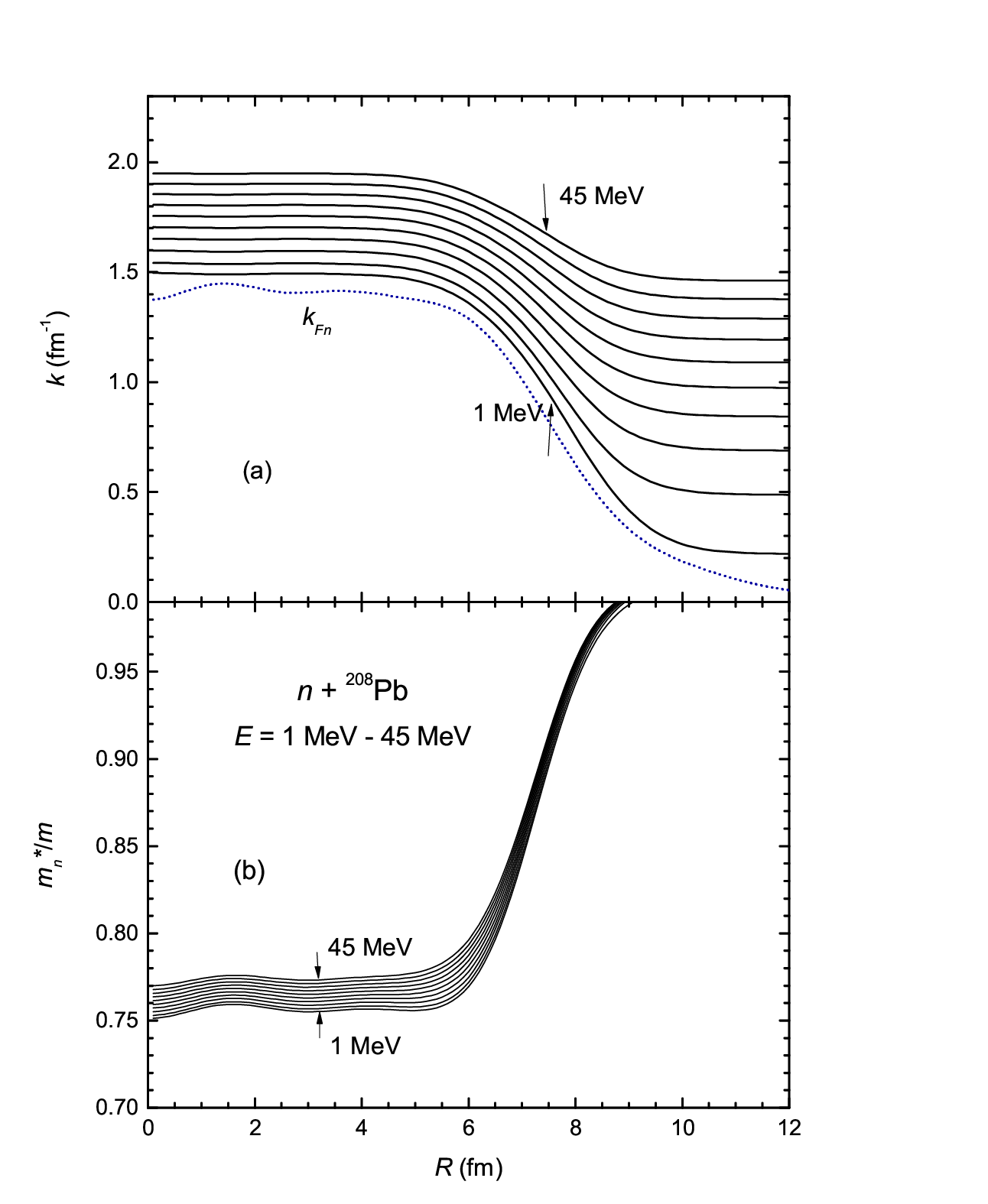}\vspace*{-0.5cm}
 \caption{(a) In-medium momentum (\ref{eq21}) of scattered neutron and the Fermi momentum $k_{Fn}$ extracted 
from the neutron ground-state density of $^{208}$Pb. (b) The radial dependence of the neutron effective 
mass (\ref{eq25}) obtained from the real folded OP at energies of 1 to 45 MeV for $^{208}$Pb target.}  \label{f8}
\end{figure}

It is obvious from Eq.~(\ref{eq21}) that the momentum $k$ of scattered nucleon depends explicitly  
on the \nA distance $R$, and the nucleon effective mass (\ref{eq25}) is, therefore, also radial dependent. 
The in-medium momentum of scattered neutron determined from the real folded $n+^{208}$Pb OP at 
energies of $E=1\sim 45$ MeV is shown in panel (a) of Fig.~\ref{f8}, and one can see that at each energy 
the neutron momentum $k$ changes gradually from its maximum of about $1.6\sim 2$ fm$^{-1}$ 
in the center to $0.2\sim 1.5$ fm$^{-1}$ at the surface, lying above the corresponding Fermi momentum. 
Over the same radial range, the neutron effective mass $m^*_n/m$ is changing from about $0.75\sim 0.78$ 
to unity at the surface. Such a radial dependence of $m^*$ is similar to that found in the nuclear structure 
studies \cite{Giai83,Lit06,Zal10}. However, the latter is usually enhanced to above unity at the surface
for the bound single-particle states lying close to the Fermi level. 
\begin{figure}[h]\vspace*{-1cm}
\includegraphics[width=0.9\textwidth]{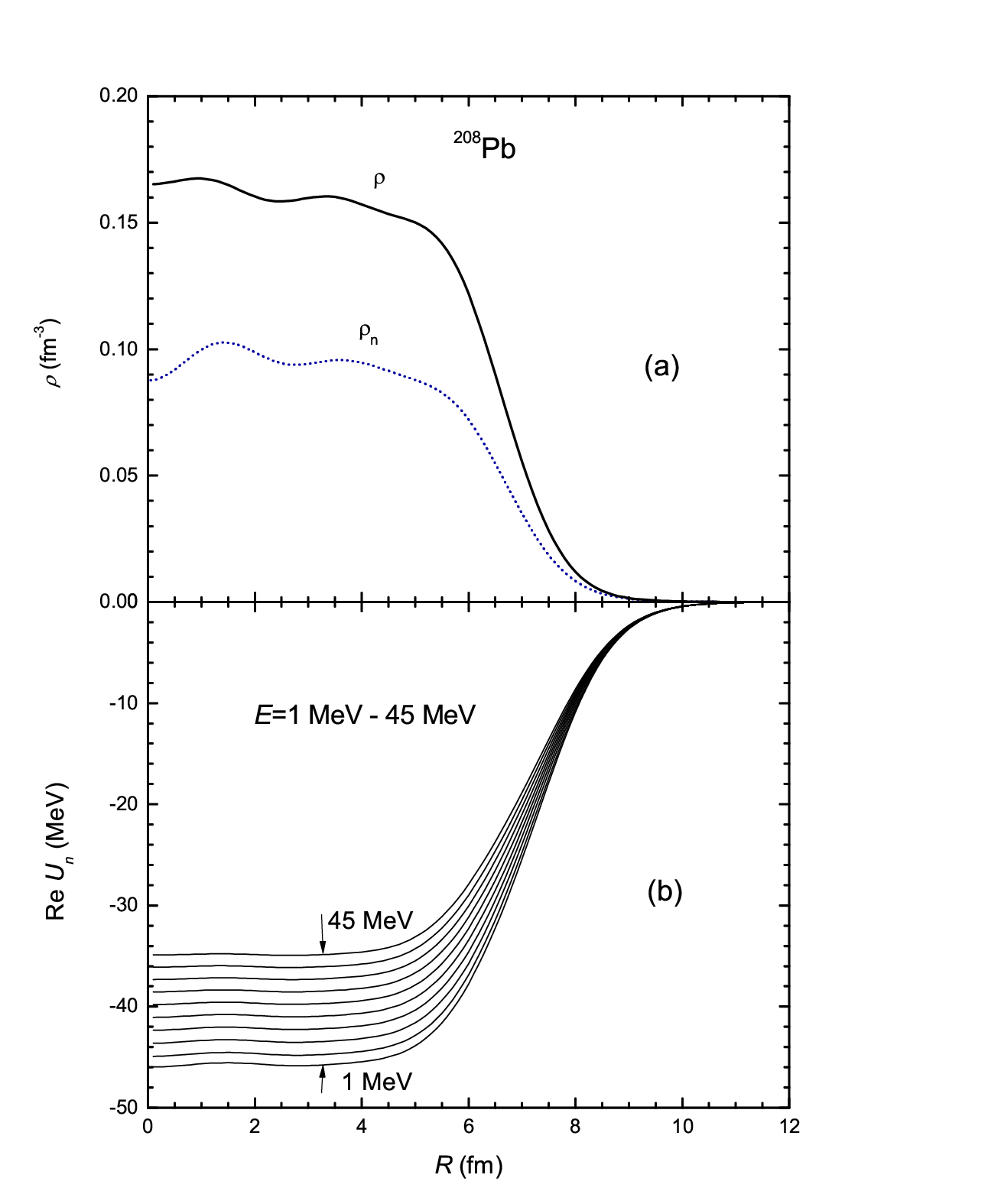}\vspace*{-0.5cm}
 \caption{(a) Neutron and total ground-state densities of $^{208}$Pb given by the HF calculation \cite{Tha11} 
 using the finite-range D1S Gogny interaction \cite{Ber91}. (b) The radial shape of the real (local) folded 
 $n+^{208}$Pb  OP obtained at energies of 1 to 45 MeV.} 
 \label{f9}
\end{figure}

\begin{figure}[h]\vspace*{-1cm}
\includegraphics[width=0.9\textwidth]{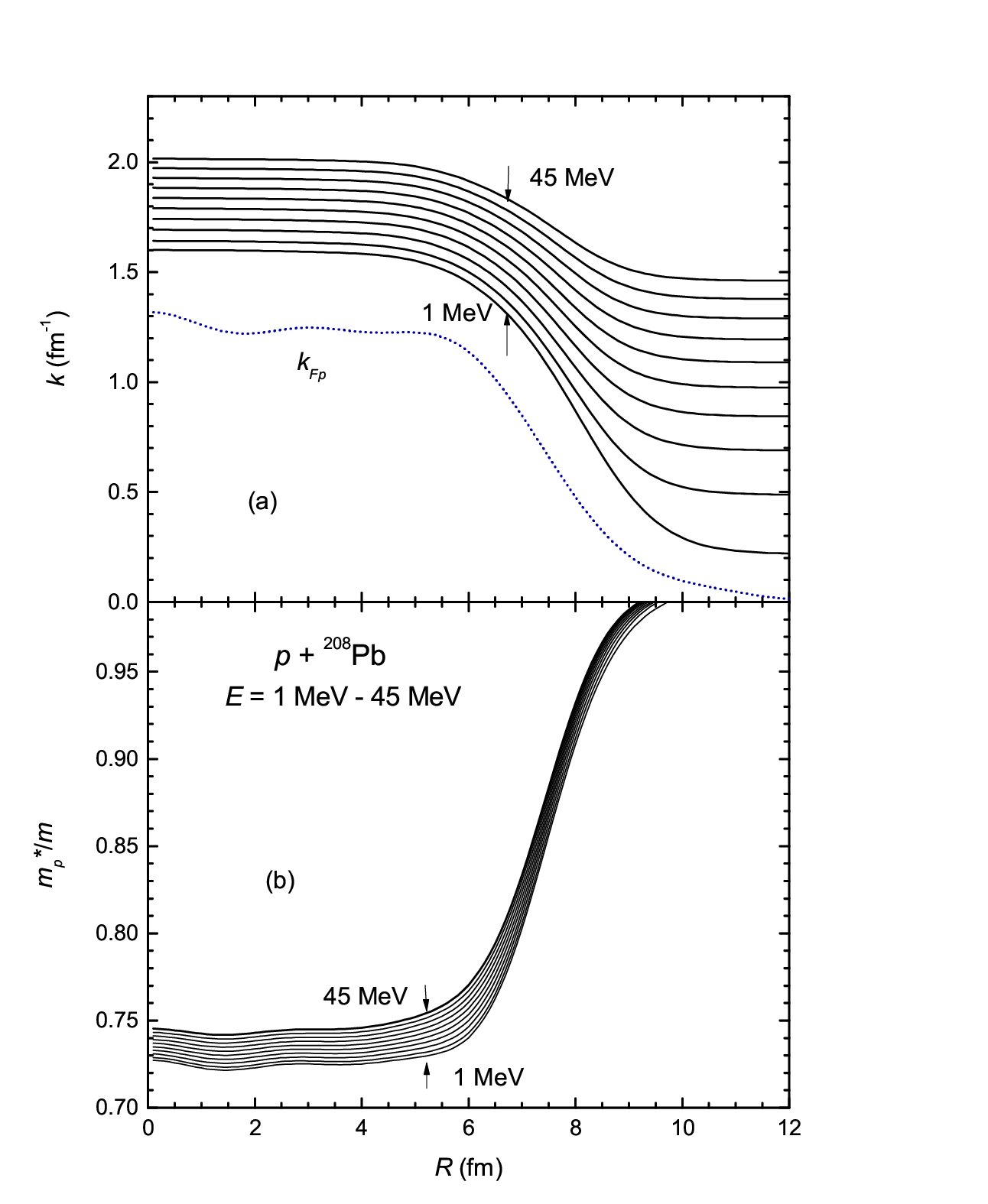}\vspace*{-0.5cm}
 \caption{The same as Fig.~\ref{f8} but for the in-medium momentum and effective mass of scattered proton.} 
\label{f10}
\end{figure}

From the radial dependence of the neutron effective mass and the neutron- and total ground-state densities 
of $^{208}$Pb shown in panel (a) of Fig.~\ref{f9}, it is straightforward to infer the density dependence 
of the effective mass of neutron scattered off $^{208}$Pb target at different energies, over the density 
range $0\lesssim\rho\lesssim\rho_0$. One can see in panel (a) of Fig.~\ref{f9} that the average 
total density in the center of $^{208}$Pb target is $\bar{\rho}\approx\rho_0$, and the neutron- and proton 
effective masses determined at distances $0\lesssim R\lesssim 3$ fm (shown in panels (b) of Figs.~\ref{f9} 
and \ref{f10}) can represent, therefore, the corresponding $m^*_\tau/m$ values in asymmetric NM at 
$\rho\approx\rho_0$, $k\gtrsim k_{F\tau}$ and neutron-proton asymmetry 
$\bar{\delta}=(\bar{\rho_n}-\bar{\rho_p})/\bar{\rho}\approx 0.185$. Because of the target valence 
neutrons, the deduced $\bar{\delta}$ value in the target center is smaller than the total difference 
in neutron- and proton numbers $(N-Z)/A$. Since our local folding model (\ref{eq13}) and (\ref{eq20}) 
predicts the real nucleon OP at $E>0$ or $k\gtrsim k_{F\tau}$, approaching the Fermi momentum from above, 
we have computed the real folded nucleon OP for $^{48}$Ca, $^{90}$Zr, and $^{208}$Pb targets at 
$E=0.05$ MeV (i.e., 50 keV above the Fermi level), and deduced the neutron- and proton effective masses 
using Eq.~(\ref{eq25}). The obtained results are presented in Table~\ref{t1} and shown in Figs.~\ref{f7} 
as rhombuses and circles. One can see that the $m^*_\tau/m$ values obtained at $\bar{\rho}\approx\rho_0$ 
and $k\gtrsim k_{F\tau}$ for these targets follow approximately the trend given by the extended HF 
calculation of the single-particle potential in NM.  
\begin{table*}
\caption{Neutron- and proton effective masses (\ref{eq25}) at the average nuclear density 
$\bar{\rho}\approx\rho_0$, deduced from the local folded nucleon OP of $^{48}$Ca, $^{90}$Zr, 
and  $^{208}$Pb targets, at distances $0\lesssim R\lesssim 3$ fm.}  
\label{t1}\vspace{0.5cm}
\begin{tabular}{|c|c|c|c|} \hline
Nucleus & $^{48}$Ca & $^{90}$Zr & $^{208}$Pb  \\ \hline
$\bar{\rho}$ & $0.159\pm 0.003$ & $0.160\pm 0.002$ & $0.160\pm 0.001$ \\ \hline
$\bar{\delta}$ & $0.0966\pm 0.0069$ & $0.0691\pm 0.0021$ & $0.1853\pm 0.0060$ \\ \hline
$m^*_n/m$  & $0.7490\pm 0.0015$ & $0.7436\pm 0.0003$ & $0.7553\pm 0.0003$ \\ \hline
$m^*_p/m$  & $0.7329\pm 0.0007$ & $0.7322\pm 0.0001$ & $0.7241\pm 0.0003$ \\ \hline
$m^*_{n-p}/\bar{\delta}$ & $0.167\pm 0.035$ & $0.165\pm 0.011$ 
 & $0.168\pm 0.009$ \\ \hline
\end{tabular}
\end{table*} 
It is interesting that the neutron-proton effective mass splitting (\ref{eq23}) obtained from the 
$m^*_\tau/m$ values given in Table~\ref{t1} also depends linearly on the asymmetry parameter, 
$m^*_{n-p}(\rho_0,\delta)\approx (0.167\pm 0.018)\delta$, which is within the empirical 
boundary of $m^*_{n-p}(\rho_0,\delta)\approx (0.27\pm 0.25)\delta$ deduced from the terrestrial 
nuclear physics experiments and astrophysical observations \cite{Li13}. In neutron-rich NM, nucleons 
at $\rho_0$ are \emph{not} bound by the in-medium NN interaction  (see, e.g., Fig~1 in Ref.~\cite{Loa15}), 
and the results obtained above for $m^*_\tau$ at $E>0$ should be appropriate for nucleons in the outer 
core of neutron star which are bound by gravitation only. 
We note further that the $m^*_{n-p}$ value estimated from the folded nucleon OP of finite nuclei 
is slightly lower than that given by nucleon OP in NM using the same interaction, 
$m^*_{n-p}\approx (0.20\pm 0.02)\delta$, and the value 
$m^*_{n-p}\approx (0.41\pm 0.15)\delta$ estimated from the phenomenological nucleon 
OP \cite{Li15}, based on the extensive OM analysis of elastic nucleon scattering. 

We note finally that the nucleon effective mass is given entirely by the momentum dependence 
of the exchange term of the folded nucleon OP, so that the nucleon effective masses presented in 
Table~\ref{t1} and shown in Fig.~\ref{f7} originated solely from the spacial Pauli nonlocality 
of nucleon OP at low energies. We have considered the energy region $0<E<50$ MeV, and found 
that the nucleon effective mass (\ref{eq25}) depends weakly on the energy. For example, 
$m^*_n/m\approx 0.7553 + 0.0004E$ and $m^*_p/m\approx 0.7241 + 0.0005E$ for $^{208}$Pb target, 
and this result agrees fairly with the empirical energy dependence of about $0.0007E$ established 
for the isoscalar $k$ effective mass of nucleons lying above the Fermi level \cite{Hog83}.   

 \section{Summary}
The generalized folding model of the nonlocal nucleon OP, with the exchange potential calculated 
exactly in the HF manner and rearrangement term properly included, has been used for the OM 
analysis of elastic neutron and proton scattering on $^{40,48}$Ca, $^{90}$Zr, and $^{208}$Pb 
targets at different energies, where the WKB local approximation for the exchange term of the 
nonlocal folded nucleon OP is validated by a consistently good OM description of the considered 
elastic data.  

Given the accurate local approximation for the nonlocal folded nucleon OP, a compact method 
is proposed to determine the nucleon effective mass at low momenta from the in-medium 
momentum dependence of the local folded nucleon OP, which originates mainly from the Pauli 
nonlocality. The results obtained for the effective mass of nucleon being scattered by the target 
mean-field potential at $0\lesssim E\lesssim 45$ MeV and $k_{F\tau}\lesssim k \lesssim 2$ fm$^{-1}$ 
seem to follow closely the isospin dependence of $m^*_\tau$ predicted by the extended HF calculation 
of single-particle potential in asymmetric NM, using the same density dependent CDM3Y6 interaction. 

The neutron-proton effective mass splitting (\ref{eq23}) given by the $m^*_\tau/m$ values 
obtained from the real folded nucleon OP of finite nuclei at low energies was found to depend 
also linearly on the asymmetry parameter $\delta$. At positive energies lying slightly above 
the Fermi level, the $m^*_{n-p}$ value determined at the center of finite nuclei (with the average
density $\bar\rho\approx\rho_0$) is slightly smaller than that obtained from the extended HF calculation 
of the single-particle potential of nucleons in asymmetric NM at $\rho_0$, where the Pauli exchange 
has been shown as the main origin of  nucleon effective mass. 

Because nucleons at density $\rho\approx \rho_0$ in the outer core of neutron star are mainly  
bound by gravitation, not by the (in-medium) density dependent NN interaction, the $m^*_\tau$ 
values obtained in this work from the folded nucleon OP of finite nuclei with neutron excess 
could be of complementary interest for the mean-field studies of the EOS of neutron star matter.    

\section*{Acknowledgments}
The authors are indebted to Pierre Descouvemont for his helpful discussions and comments 
on the calculable $R$-matrix method \cite{desco,desco2}. We also thank Bao-An Li for his 
communication and suggestion on the nucleon effective mass. The present research has been 
supported, in part, by the National Foundation for Scientific and Technological Development 
of Vietnam (NAFOSTED Project No. 103.04-2021.74).

\end{document}